# Superconductivity in Zirconium Polyhydrides with Tc above 70K

Changling Zhang [1,2], Xin He [1,2,3], Zhiwen Li [1,2], Sijia Zhang [1], Shaomin Feng [1], Xiancheng Wang [1,2], Richeng Yu [1,2], Changqing Jin [1,2,3]

[1] Beijing National Laboratory for Condensed Matter Physics, Institute of Physics, Chinese Academy of Sciences, Beijing 100190, China

[2] School of Physical Sciences, University of Chinese Academy of Sciences, Beijing 100190, China

[3] Songshan Lake Materials Laboratory, Dongguan 523808, China

Hydrogen is expected to be a metal at sufficient high pressures[1] while it can also be a high temperature superconductor based on BCS theory because of its high Debye temperature arising from the light mass of hydrogen[2]. Since it is very challenge to reach the pressure for the hydrogen metallization, the "chemically precompressed" hydrogen rich compounds were in turn proposed[3, 4] to realize high temperature superconductivity at lower pressures. The experimental discovery of high temperature superconductivity[5] in $SH_3$ has greatly stimulated the efforts to exploring new hydrogen rich superconductors such as in rare earth polyhydrides[6, 7]. Here we report discovery of superconductivity with Tc about 71 K in zirconium polyhydrides, the record high Tc so far for *4d* transition metal hydrides. The new materials are synthesized above 200 GPa & 2000 K using diamond anvil cell (DAC) combined with laser heating.

The diamond anvil cells made of CuBe with 30 over 300 μm beveled culet are used to produce high pressure up to megabar level. A plate of T301 stainless steel covered with cBN powder was used as a gasket that was indented to 10 μm thickness. A hole of approximately 20 μm in diameter was drilled in the indented gasket to serve as the sample chamber. A thin square shaped zirconium specimen with size of 8 μm*8 μm*2 μm locates at the center of culet on which four Pt electrodes are sputtered for the transport measurements[8]. The ammonia borane (AB) was filled in the chamber as both hydrogen source (hydrodizer) and the pressure transmitting medium[6]. The samples are pressed to 210 to 240 GPa followed by laser heating to 2000 K for a few minutes. The heating was performed with a YAG laser of 1064 nm wave length in a continuous mode. The temperature was estimated by fitting black body irradiation spectra. The sample was quenched upon laser heating but maintained at the same pressure for transport measurements in a Maglab system combined with diamond anvil cell that provides the synergetic environments of high pressure at megabar order, low temperature down to 1.5 K and high magnetic field up to 9 T.

**Results and discussions**

Fig. 1(a) displays the temperature dependence of resistance measured at 220 GPa for a zirconium polyhydride sample. The resistance shows a rapid drop at ~71 K and approaches to zero with temperature decrease that suggests superconducting transition. The inset of Fig. 1(a) is the derivative of resistance over temperature. The $T_c^{onset}$ & $T_c^{zero}$ are ~71 K and 63 K that are determined by the right & left side upturn point, respectively. Calculations suggested that zirconium polyhydrides might be formed at high pressures with possible high temperature superconductivity[9] but only $T_c$~6.4 K was experimentally reported so far[10]. It is theoretically predicted that the $Cmc2_1$ ZrH$_6$ is stable within the pressure of 150 ~ 275 GPa and $T_c$ is estimated to be 70 K at 215 GPa [9] that in principle matches with our experimental results in terms of pressure and $T_c$. Hence the observed superconductivity here might arise from the $Cmc2_1$ ZrH$_6$. Since AB was used as the hydrogen source, we cannot rule out the possibility of N or B doping into the synthesized zirconium polyhydrides, which might have effect on $T_c$ [11]. Fig. 1(b) shows the resistance versus temperature measured at applied magnetic fields. The transition temperature shifts gradually to low temperature with increasing field, in consistent to the nature of superconducting transitions. Fig. 1(c) presents the results for another sample synthesized at 235 GPa, showing transition with the same $T_c$ while the transition almost coincides in both warming & cooling cycles.

Previously there are a number of reports on high $T_c$ superconductivity in hydrides including of sulfur[5, 12], rare-earth[6, 7, 13], alkaline-earth[14, 15] etc. The discovery of high $T_c$ superconductivity in zirconium polyhydrides sheds light to search for new superconductors in transition metal hydrides.

**Acknowledgments:**

The work is supported by NSF, MOST & CAS of China through research projects. We are grateful to Prof. J. G. Cheng, J.P.Hu & L. Yu for the discussions. We thank Prof. T. Xiang, B. G. Shen, & Z. X. Zhao for the consistent encouragements.

**Author Contributions**

Research Design & Supervision: Changqing Jin; high pressure synthesis and in situ resistance measurements: Changling Zhang, Xin He, Zhiwen Li, Sijia Zhang, Shaomin Feng, Xiancheng Wang, Richeng Yu, Changqing Jin; manuscript writing: Changling Zhang, Xiancheng Wang & Changqing Jin. All authors contributed to the discussions.

**Figure Captions:**

(a) The temperature dependence of resistance measured at 220 GPa for a zirconium polyhydride synthesized at the same pressure at 2000 K. The inset is the derivative of resistance over temperature to determine the critical transition temperature.

(b) The resistance measured under applied magnetic fields for the sample.

(c) The resistance for another sample measured at 235 GPa. The inset is the change with applied magnetic fields.

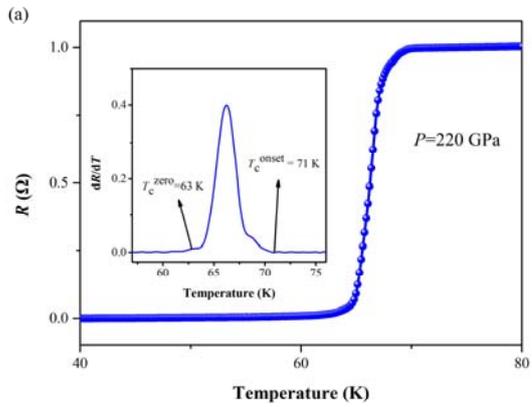

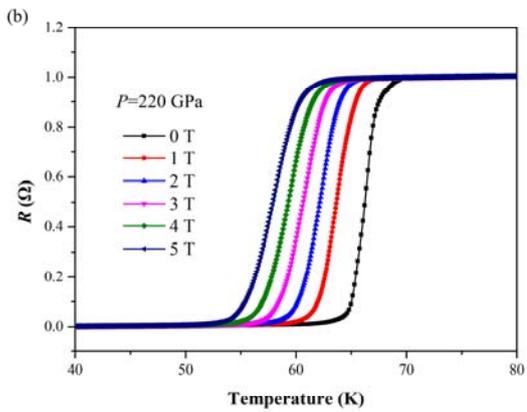

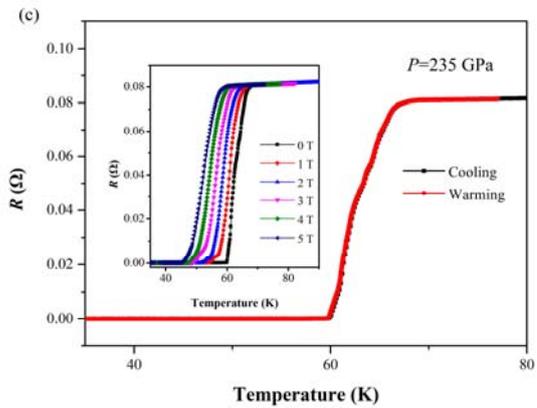